\documentclass[11pt,twoside]{article}

\usepackage{asp2004}
\usepackage{psfig}
\usepackage{epsf}
\usepackage{graphics}
\usepackage{lscape}

\def\edcomment#1{\iffalse\marginpar{\raggedright\sl#1\/}\else\relax\fi}
\reversemarginpar

\begin{document}

\title{%
Chandra HETG Spectra of SS Cyg and U Gem in Quiescence and Outburst}

\author{%
Christopher W.\ Mauche$^1$,
Peter J.\ Wheatley$^2$,
Knox S.\ Long$^3$, \\
John C.\ Raymond$^4$, and
Paula Szkody$^5$}

\affil{%
$^1$Lawrence Livermore Nat'l Lab., 7000 East Ave., Livermore, CA, USA\\
$^2$University of Leicester, Dept.\ of Physics \& Astronomy, Leicester, UK\\
$^3$STScI, 3700 San Martin Drive, Baltimore, MD, USA\\
$^4$Center for Astrophysics, 60 Garden Street, Cambridge, MA, USA\\
$^5$University of Washington, Dept.\ of Astronomy, Seattle, WA, USA}

\begin{abstract}
{\it Chandra\/} HETG spectra of the prototypical dwarf novae SS~Cyg
and U~Gem in quiescence and outburst are presented and discussed.
When SS~Cyg goes into outburst, it becomes dimmer in hard X-rays and
displays a dramatic shift in its relative line strengths, whereas when
U~Gem goes into outburst, it becomes brighter in hard X-rays and
displays only a minor shift in its relative line strengths. In both
systems, the emission lines become significantly broader in outburst,
signaling the presence of high velocity gas either in Keplerian orbits
around the white dwarf or flowing outward from the system.
\end{abstract}

The X-ray spectra of nonmagnetic cataclysmic variables provide important
and unique information about the nature (the mass-accretion rate,
emission measure distribution, density, velocity, abundances) of the
boundary layer between the accretion disk and the surface of the white
dwarf, where (nominally) half of the gravitational potential energy of
the accreted material is released. The response of the boundary layer
to changes in the mass-accretion rate is best studied in dwarf novae,
where the mass-accretion rate varies systematically by a factor of
$\sim 10^3$ between quiescence and outburst. Toward this end, we
observed the prototypical dwarf novae SS~Cyg and U~Gem in quiescence
and outburst with the {\it Chandra\/} X-ray Observatory High Energy
Transmission Grating Spectrometer (HETGS). {\it Chandra\/} observations
were obtained of SS~Cyg in quiescence on 2000 August 24 and near the
peak and early decline of a narrow outburst on 2000 September 12 and
14 (exposures of 47 and 96 ks), and of U~Gem in quiescence on 2000
November 29 and at the peak of outburst on 2002 December 26 (exposures
of 95 and 61 ks). The quiescent spectra of both systems are discussed
briefly by \cite{muk03}, while a detailed account of the quiescent
spectrum of U~Gem is provided by \cite{szk02}. In this brief
communication, we show the quiescent and outburst spectra of both
systems in Figure 1 to highlight their similarities and differences.

In all cases, the X-ray spectra contain emission lines of a broad
range of ions, consistent with the spectra of multi-temperature thermal
plasmas. When SS Cyg goes into outburst, it becomes dimmer in hard
X-rays and displays a dramatic shift in its relative line strengths
(from predominately H-like to predominately He-like lines), whereas
when U Gem goes into outburst, it becomes brighter in hard X-rays and
displays only a minor shift in its relative line strengths. In both
cases, the emission lines become significantly broader in outburst,
signaling the presence of high velocity gas either in Keplerian orbits
around the white dwarf or flowing outward from the system. Such a wind
is seen (in lower excitation lines) in the {\it Chandra\/} Low Energy
Transmission Grating spectrum of SS~Cyg in outburst \citep{mau04}.

\begin{figure}[!ht]
\plotone{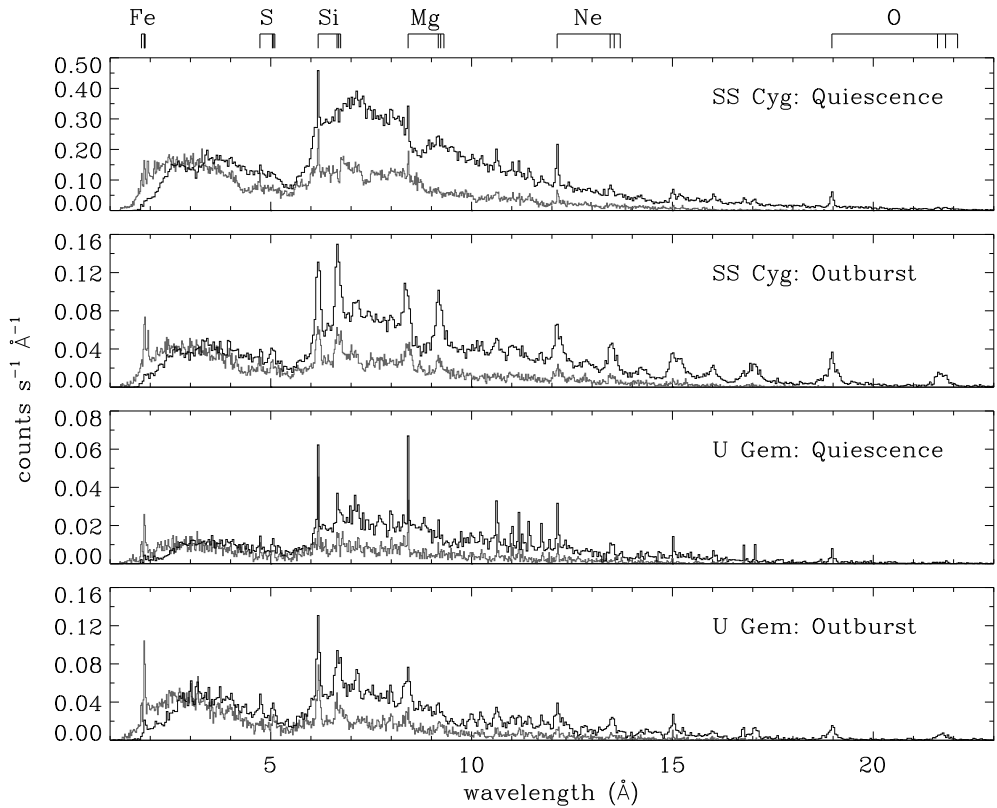}
\vskip -10pt
\caption{{\it Chandra\/} MEG ({\it upper histograms\/}) and HEG ({\it 
lower histograms\/}) count spectra of SS~Cyg and U~Gem in quiescence
and outburst, with identifi\-cations of emission lines of H- and He-like
ions of O, Ne, Mg, Si, S, and Fe.}
\end{figure}

\acknowledgments

Support for this work was provided in part by NASA through
{\it Chandra\/} Award Numbers GO0-1094A and GO3-4025B issued by the
{\it Chandra\/} X-ray Observatory Center, which is operated by SAO for
and on behalf of NASA under contract NAS8-03060. This work was performed
under the auspices of the US Department of Energy by University of
California Lawrence Livermore National Laboratory under contract
W-7405-Eng-48.

\end{document}